\documentclass[showpacs, twocolumn]{revtex4-1}
\usepackage{graphicx}
\usepackage{amsmath}
\usepackage{appendix}

\begin{document}

\title{Quantum self-bound droplets in Bose-Bose mixtures: Effects of higher-order quantum and thermal fluctuations}

\author{Nadia Guebli $^1$ and Abdel\^{a}ali Boudjem\^{a}a $^{2,3}$ }
\affiliation{$^1$ Laboratory for Theoretical Phyiscs and Material Physics, Hassiba Benbouali University of Chlef, P.O. Box 78, 02000, Ouled-Fares, Chlef, Algeria.\\
$^2$ Department of Physics, Faculty of Exact Sciences and Informatics, Hassiba Benbouali University of Chlef, P.O. Box 78, 02000, Ouled-Fares, Chlef, Algeria.\\
$^3$Laboratory of Mechanics and Energy, Hassiba Benbouali University of Chlef, P.O. Box 78, 02000, Ouled-Fares, Chlef, Algeria.}
\email {a.boudjemaa@univ-chlef.dz}

\date{\today}

\begin{abstract}

We systematically study the effects of higher-order quantum and thermal fluctuations on the stabilization of self-bound droplets in Bose mixtures
employing the time-dependent Hartree-Fock-Bogoliubov theory.
We calculate the ground-state energy, the droplet equilibrium density, the depletion and anomalous density of the droplet as well as the critical temperature
as a function of the relevant parameters. 
Our findings are compared with previous analytical predictions and diffusion Monte Carlo simulations.
We employ our theory together with the local density approximation for quantum and thermal fluctuations  to obtain an extended finite-temperature Gross-Pitaevskii equation. 
The density profiles and breathing modes of the droplet are deeply examined in terms of the interaction strength and the temperature by numerically solving 
the developed generalized Gross-Pitaevskii equation.

\end{abstract}

\maketitle

\section{Introduction} \label{Intro}

Recently, quantum self-bound droplets in Bose-Bose mixtures have attracted significant theoretical and experimental interest \cite{Petrov,Capp,Cik, Cab,Sem,Err}.
This exotic state of matter could pave the way to investigate beyond-mean-field intriguing quantum phenomena.
It arises from the competition of an attractive mean-field energy and repulsive beyond-mean-field effects provided by the  
Lee-Huang-Yang (LHY) term. This latter stems from the quantum fluctuations induced by interactions.
Self-bound droplets in Bose mixtures have been the object of several recent works.
The role played by  dipole-dipole interactions in the stability of the droplets in binary Bose-Einstein condensates (BECs) was first highlighted by one of us \cite{Boudj}.
The effects of correlations in self-bound Bose mixtures have been explored using the variational hypernetted-chain Euler-Lagrange method \cite{Stau}.
The dynamical properties of low-dimensional self-bound quantum droplets of bosonic mixtures have been studied in Refs.\cite{Astr,Chiq, Astr1,Astr2,Astr3} 
by solving the corresponding amended Gross-Pitaevskii equation (GPE).
Most recently, the self-bound supersolid stripe phase in binary BEC has been predicted in Ref.\cite{Sach}.

        
The aim of this paper is to investigate the effects of quantum and thermal fluctuations on the occurrence and stabilization of quantum droplets in symmetric Bose mixtures
using the beyond LHY description based on our time dependent Hartree-Fock-Bogoliubov (TDHFB) theory \cite{Boudj, Boudj1, Boudj2,Boudj3, Boudj4,
 Boudj5, Boudj6, Boudj7, Boudj8,  Boudj9, Boudj10, Boudj11, Boudj12, Boudj13}.  
This latter includes automatically higher-order corrections originating from normal and anomalous correlations to the equation of state generalizing all existing models in the literature. 
The coupling between the order parameter, and the normal and anomalous fluctuations makes the TDHFB  a promising approach  for exploring quantum 
self-bound droplets in a dipolar single BEC \cite{Boudj12} and Bose-Bose mixtures \cite{Boudj13} at both zero and finite temperatures.

In this paper we extract useful analytical results for the ground-state energy and the equilibrium density
that extend naturally the seminal equations of Petrov \cite{Petrov}.
Explicit expressions connecting normal and anomalous correlations to the droplet equilibrium density are also obtained for the first time to the best of our knowledge.
We show that at zero temperature, our theory not only provides fascinating results but also captures genuine higher-order quantum effects 
predicted from recent diffusion Monte Carlo (DMC) simulations \cite{Cik1}. 
At finite temperature, we calculate the free energy and precisely determine the thermal equilibrium density of the droplet.
We find that the droplet destabilizes when the temperature becomes slightly larger than the ground-state energy of the droplet,
in good agreement with recent results inferred from the macroscopic approach \cite{Ota} and the pairing theory \cite{Hu}.
Our predictions point out that higher-order corrections and interspecies interactions may shift the critical temperature. 

On the other hand, we  deal with the ground-state properties of the droplet such as the density profiles and the collective modes. 
These latter play a crucial role for probing many-body effects in ultracold quantum gases.
Our calculations are based on the extended finite-temperature GPE which is derived in a self-consistent manner from our formalism
taking into account higher-order quantum and thermal corrections under the local density approximation.  
Our goal here is to understand how the density distribution and the collective oscillations behave when varying the interaction strength and temperature.
To this end, we solve numerically and variationally the developed GPE.

The rest of the paper is organized as follows. In Sec.\ref{Mod}, we recapitulate the TDHFB theory.
We briefly review the expressions of the free energy, and the normal and anomalous fluctuations.
In Sec.\ref{SBD}, we present  our calculations up to second-order terms for the energy per particle and equilibrium density. 
Our results are compared with available theoretical treatments and DMC simulations.
The behavior of the noncondensed and anomalous densities of the droplet is also highlighted.
Furthermore, we look at how the thermal fluctuations influence the nucleation and stability of the self-bound droplet. 
We compute also the critical temperature and thermal equilibrium density of the self-bound droplet.
Section \ref{GSP} is devoted to the numerical analysis of the ground-state properties of the self-bound droplet. 
We calculate in particular diverse density profiles and the collective modes.
The paper ends with conclusions and an outlook in Sec.\ref{concl}.




\section {Model} \label{Mod}

We consider a weakly interacting two-component BEC  with an atomic mass $m_j$.
In the mean field, the physics of this system is given by the following energy functional which includes the LHY quantum corrections and normal and anomalous densities \cite{Boudj,Boudj5,Boudj13},
\begin{align}  \label{egy}
{\cal E} &= \sum_{j=1}^2 \bigg[ \int d{\bf r} \, \left( \Phi_j^*  h_j^{sp} \Phi_j + \hat{\bar \psi}_j^\dagger  h_j^{sp}  \hat{\bar \psi}_j +\frac{g_j}{2}  n_j^2 \right) \bigg] \\
&+ g_{12} \int d{\bf r}  n_1n_2 + {\cal E}_{\text {LHY}}, \nonumber
\end{align}
where $h_j^{sp} =-(\hbar^2 /2m_j) \Delta -\mu_j$ is the single particle Hamiltonian, $\mu_j$ represent the chemical potentials related to each component, 
$n_{cj}=|\Phi_j|^2$ is the condensed density, and
$\hat{\bar \psi}_j({\bf r})=\hat\psi_j({\bf r})- \Phi_j({\bf r})$ is the noncondensed part of the field operator with $\Phi_j({\bf r})=\langle\hat\psi_j({\bf r})\rangle$. 
The coefficients $g_j=(4\pi \hbar^2/m_j) a_j$ and $g_{12}=g_{21}= 2\pi \hbar^2 (m_1^{-1}+m_2^{-1}) a_{12}$ with 
$a_j$ and $a_{12}$ being the intraspecies and the interspecies scattering lengths, respectively.
In the spirit of our HFB theory,  the LHY corrections to the energy are defined as:
\begin{align} \label{LHY}
{\cal E}_{\text{LHY}} &= \frac{1}{2}\sum_{j=1}^2 g_j \int d{\bf r} \big( 2\tilde n_j n_j-\tilde n_j^2 +|\tilde m_j|^2 \\
&+ \tilde m_j^*\Phi_j^2+ \tilde m_j {\Phi_j^*}^2 \big), \nonumber
\end{align}
where $\tilde n_j=\langle\hat{\bar {\psi}}_j ^\dagger\hat{\bar {\psi}}_j\rangle$ is the noncondensed density, 
and $\tilde m_j= \langle\hat {\bar {\psi}}_j\hat{\bar {\psi}}_j\rangle$ is the anomalous density. 
The presence of these quantities enables us to derive the LHY term without any {\it ad-hoc} assumptions in contrast to the standard generalized GPE.


Let us start by briefly reviewing the main results predicted by our theory \cite{Boudj13} that permit us to study the properties of the self-bound droplets in Bose mixtures.

The free energy can be computed by resorting to the dimensional regularization. This yields the famous $T^4$-law \cite{Boudj13}
\begin{align}\label{LHY1}
\frac{F}{V} =& \frac{1}{2} \sum_{j=1}^2 g_j n_j^2 + g_{12}  n_1 n_2 +\frac{16  g_1 \sqrt{a_1^3/\pi}}{15\sqrt{2}} n_1^{5/2} \nonumber\\
&\times \bigg(1+\frac{\tilde m_1-\tilde n_1}{n_1}\bigg)^{5/2} \Bigg\{ \sum_{\pm} \bigg[ f_{\pm}^{5/2}(\Delta,\alpha)  \nonumber \\
&-\frac{1}{2\sqrt{2}}  \frac{f_{\pm}^{-3/2}(\Delta,\alpha) }{[1+(\tilde m_1-\tilde n_1)/n_1]^4}\left(\frac{\pi T}{ g_1 n_1 }\right)^4\bigg] \Bigg\}, 
\end{align}
where  $V$ is the volume of the system, $f_{\pm} (\Delta, \alpha)= 1 + \alpha \pm \sqrt{ (1-\alpha)^2 +4 \Delta ^{-1}\alpha }$, $\alpha= \bar g_2 n_2/ \bar g_1 n_1$,
and $\Delta=\bar g_1\bar g_2/g_{12}^2$ is the miscibility parameter.
The last term which accounts for the effect of the thermal fluctuations, is calculated at temperatures $T\ll g_1n_1$, where the main contribution to Eq.(\ref{LHY1}) comes
from the phonon region.
In the limit of lower density $n \rightarrow 0$, the thermal corrections to the free energy  diverge as $n^{-3/2}$  and vanish at zero temperature.
A similar formula has been derived in Ref.\cite{Ota} using the beyond-LHY theory, based on the calculation of second-order terms in the Bogoliubov phonon modes.
At $T=0$ and for $\tilde m_j=\tilde n_j=0$, the free energy (\ref{LHY1}) reproduces the analytical expression for the ground-state energy \cite{Petrov}.

Our formalism provides useful expressions for the noncondensed and anomalous densities at finite temperature \cite{Boudj13},
\begin{align}\label {NCD}
\tilde n &= \frac{2\sqrt{2} }{3} \,n_{1} \sqrt{ \frac{n_{1} a_1^3}{\pi}} \bigg(1+\frac{\tilde m_1-\tilde n_1}{n_1}\bigg)^{3/2} \\
&\times\sum_{\pm} \bigg[ f_{\pm}^{3/2} (\Delta,\alpha) +\frac{1 }{3}\frac{f_{\pm}^{-1/2}(\Delta,\alpha) }{[1+(\tilde m_1-\tilde n_1)/n_1]^2}
\left(\frac{\pi T}{ g_1 n_1 }\right)^2\bigg],\nonumber
\end{align}
and 
\begin{align}\label {MCD}
\tilde m &= 2\sqrt{2} \,n_{1}  \sqrt{ \frac{n_{1} a_1^3}{\pi}} \bigg(1+\frac{\tilde m_1-\tilde n_1}{n_1}\bigg)^{3/2} \\
&\times\sum_{\pm} \bigg[ f_{\pm}^{3/2}(\Delta,\alpha)+\frac{1 }{3} \frac{f_{\pm}^{-1/2}(\Delta,\alpha) } {[1+(\tilde m_1-\tilde n_1)/n_1]^2}
\left(\frac{\pi T}{ g_1 n_1 }\right)^2\bigg].\nonumber
\end{align}
For $g_{12}=0$, Eqs.(\ref{NCD}) and (\ref{MCD}) reduce to those obtained in our Ref.\cite{Boudj15} for a single component BEC.
The expression of the anomalous density (\ref{MCD}) has been indeed obtained using the dimensional regularization \cite{Boudj3,Boudj12,Anders} to overcome the ultraviolet divergence.
It is worth stressing that the involvement of the anomalous density in Bose systems leads to a double counting of the interaction effects \cite{Boudj2, Boudj10,Griffin,DStoof}.
The solution of the self-consistent Eqs.(\ref{NCD}) and (\ref{MCD}) requires the use of some form of an iterative scheme.

\begin{figure}
\centerline{
		\includegraphics[width=0.45\linewidth]{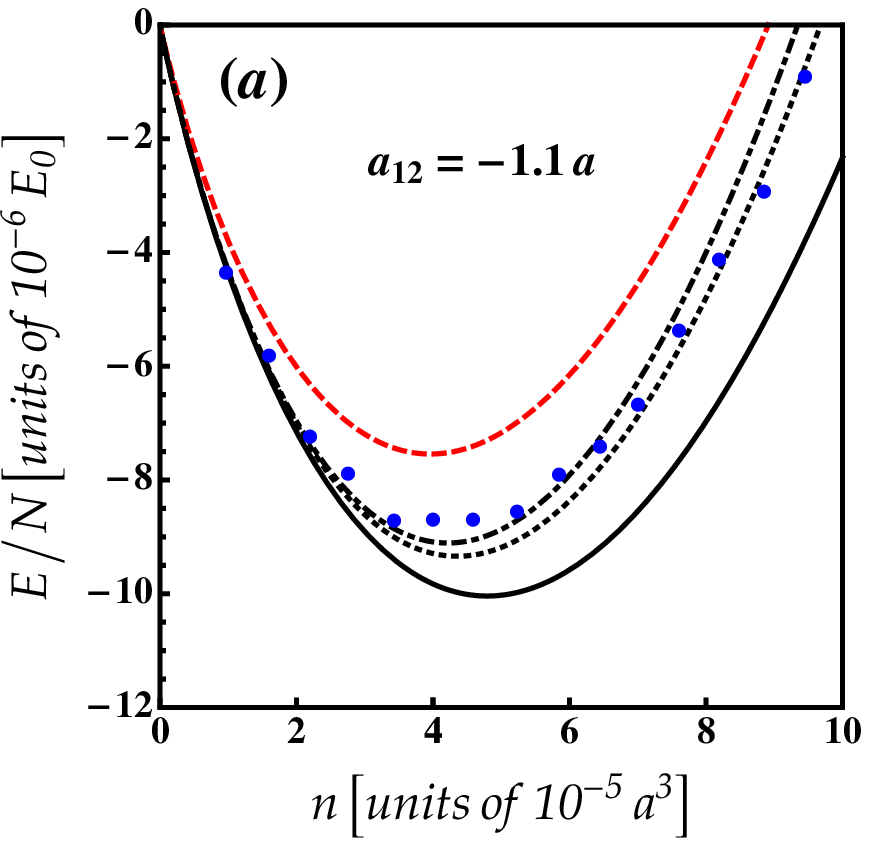}
		\includegraphics[width=0.45\linewidth]{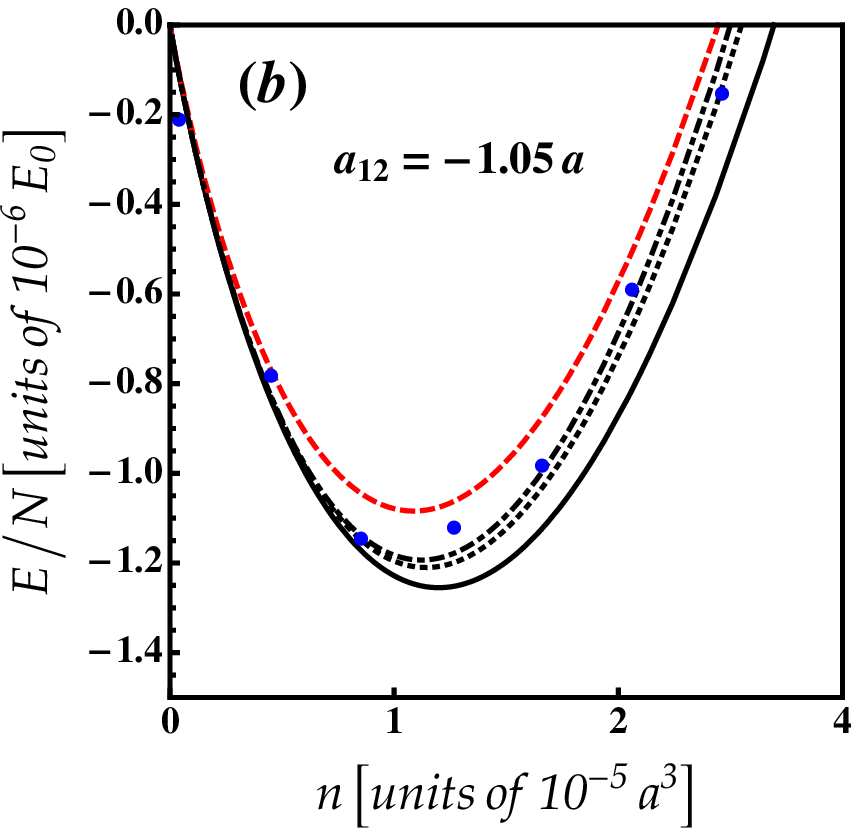}}
		\caption{(Color online)  (a) Ground state energy as a function of the density for $a_{12}/a=-1.1$. (b) The same as (a) but for $a_{12}/a=-1.05$.
Red dashed lines correspond to the pairing theory \cite{Hu}.
Black dotted-dashed lines correspond to our results to  first-order corrections of quantum fluctuations. 
Black dotted lines correspond to our results up to second-order corrections of quantum fluctuations.
Black solid lines correspond to Petrov's predictions \cite{Petrov}. Circles represent the DMC results of Ref.\cite{Cik1}.}
\label{egy} 
\end{figure}



\section{Self-bound droplet} \label{SBD}

Here, we apply the general method presented in the previous section to investigate the properties of the self-bound droplet at both zero and finite temperatures
in the presence of higher-order quantum effects.

From now on, we consider symmetric Bose-Bose mixtures with equal intraspecies coupling constants $g_1=g_2=g$, and equal densities in each component $n_1=n_2=n/2$, $\tilde m_1=\tilde m_2=\tilde m/2$, and $\tilde n_1=\tilde n_2=\tilde n/2$.

\subsection{Zero-temperature case}


\begin{figure} 
\centerline{
\includegraphics[width=0.47\linewidth]{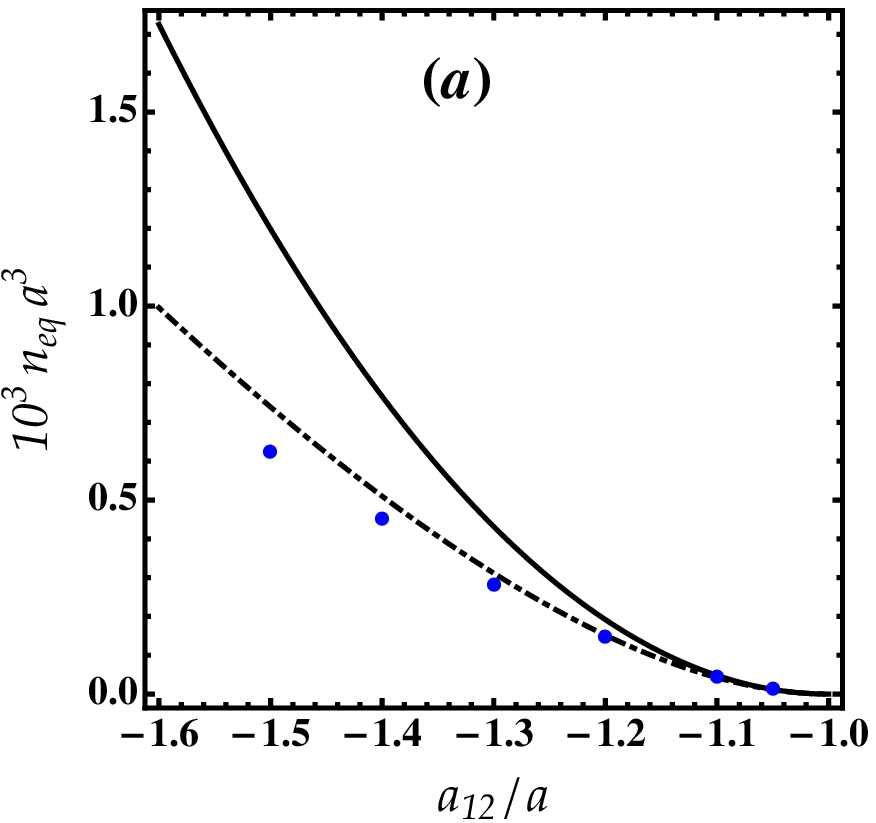}
\includegraphics[width=0.47\linewidth]{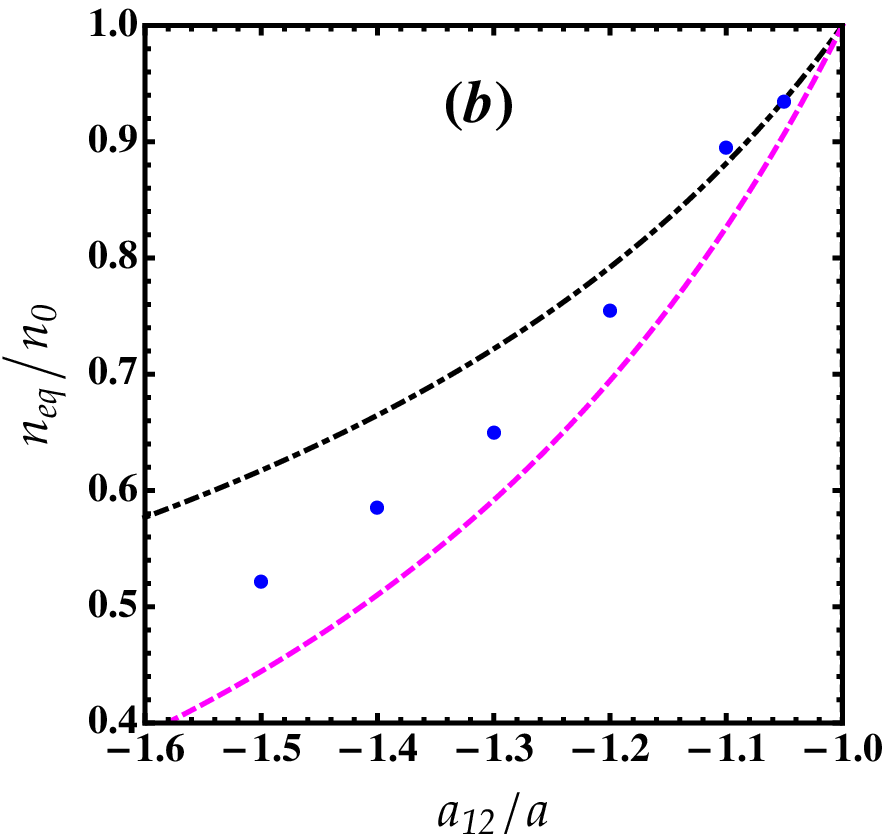}}
\caption{ (a) Equilibrium density of the droplet as a function of $a_{12}/a$. 
(b) Equilibrium density in units of $n_0$, as a function of $a_{12}/a$.
Black solid lines correspond to Petrov's equilibrium density \cite{Petrov}. 
Black dotted-dashed lines correspond to our results from Eq. (\ref{eqd}). Magenta dashed lines correspond to the pairing theory of Ref.\cite{Hu}. Blue circles correspond to DMC data \cite{Cik1}.}
\label{equi} 
\end{figure}

The ground state energy can be rewritten in the following dimensionless form:
\begin{align}\label{LHY2}
\frac{E}{NE_0} =& \pi \left(\frac{\delta a}{a}\right)_+ (na^3)+\frac{32 \sqrt{\pi}}{15} (n a^3)^{3/2} \nonumber\\
&\times \bigg(1+\frac{\tilde m-\tilde n}{n}\bigg)^{5/2} \sum_{\pm}  \left(\frac{\delta a}{a}\right)_{\pm}^{5/2}, 
\end{align}
where $E_0=\hbar^2/ma^2$, and $(\delta a/a)_{\pm}=1\pm (a_{12}/a)$.
Equation (\ref{LHY2}) extends and unifies various equations of state in the existing literature.
For $\left(\delta a/a\right)_+ <0$, the mean-field theory energy provides a term $\propto n^2$ leading to a collapse of a homogeneous state towards bright soliton formation.
The beyond-mean-field LHY term adds an extra repulsive term  $\propto n^{5/2}$ compensating
the attractive mean-field term, enabling quantum and thermal fluctuations to stabilize mixture droplets against collapse.

We solve iteratively the energy (\ref{LHY2}) up to second-order in $\tilde n$ and $\tilde m$. 
Our results are compared with recent DMC data \cite{Cik1} and theoretical analyses  \cite{Petrov, Hu}. 
Figure \ref{egy} shows that our findings are in excellent agreement with the DMC simulations \cite{Cik1}, 
and improve upon the conventional Bogoliubov and the pairing predictions previously reported in \cite{Petrov, Hu} regardless o the value of $a_{12}$. 
This indicates the importance of the beyond-LHY quantum corrections.

\begin{figure} [hp]
\centerline{
		\includegraphics[width=0.8\linewidth]{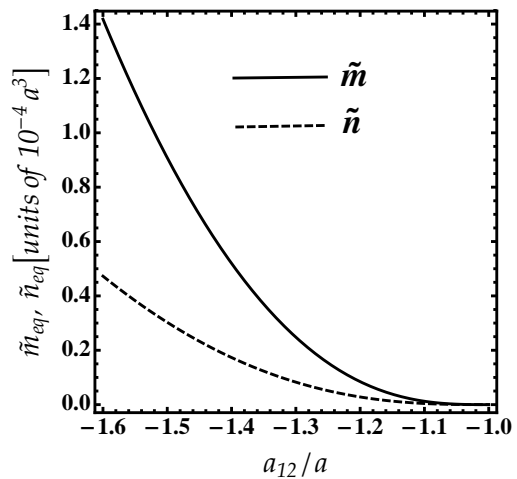}}
		\caption{ Anomalous and noncondensed equilibrium density of the self-bound droplet state from Eqs.(\ref{ncdeq}) and (\ref{andeq}) as a function of $a_{12}/a$.}
\label{AND} 
\end{figure}

The equilibrium density of the droplet can be obtained by minimizing the energy with respect to the density  \cite{Petrov}.
Using the identity $(1+a_{12}/a)^{j/2}+(1-a_{12}/a)^{j/2}=2^{j/2}$, we immediately get
\begin{align}  \label{eqd}
 \frac{n_{eq}}{n_0}&=\frac{0.0576}{\left(\delta a/a\right)_+^2} \Big[18 -25 \left(\delta a/a\right)_+ \\
&-6\sqrt{9  -25 \left(\delta a/a\right)_+}\Big]. \nonumber
\end{align}
This equation differs by the term $[0.0576/\left(\delta a_+/a\right)^2] [18 -25 \left(\delta a_+/a\right)
-6\sqrt{9  -25 \left(\delta a_+/a\right)}]$ from the equilibrium density $n_0a^3=\left(25 \pi/16384 \right) \left(\delta a_+/a\right)^2$ 
predicted by Petrov \cite{Petrov}. 
In Fig.\ref{equi}  we plot our predictions for the equilibrium density and compare the results with the DMC simulation points  \cite{Cik1} and the 
existing previous theoretical models  \cite{Petrov,Hu}.
Figure \ref{equi}.(a) shows that our theoretical results based on the HFB theory  are in good agreement with the DMC simulation practically in the whole range of interactions.
However, for large values of $|a_{12}/a|$,  the ratio $n_{eq}/n_0$ assessed by our theory diverges from that of the Bogoliubov  approach \cite{Petrov} and the pairing theory \cite{Hu} 
[see Fig.\ref{equi}.(b)].
One might argue that quantum fluctuations are responsible for such an increase in our equilibrium density.

\begin{figure}
	\centerline{
		\includegraphics[scale=0.47]{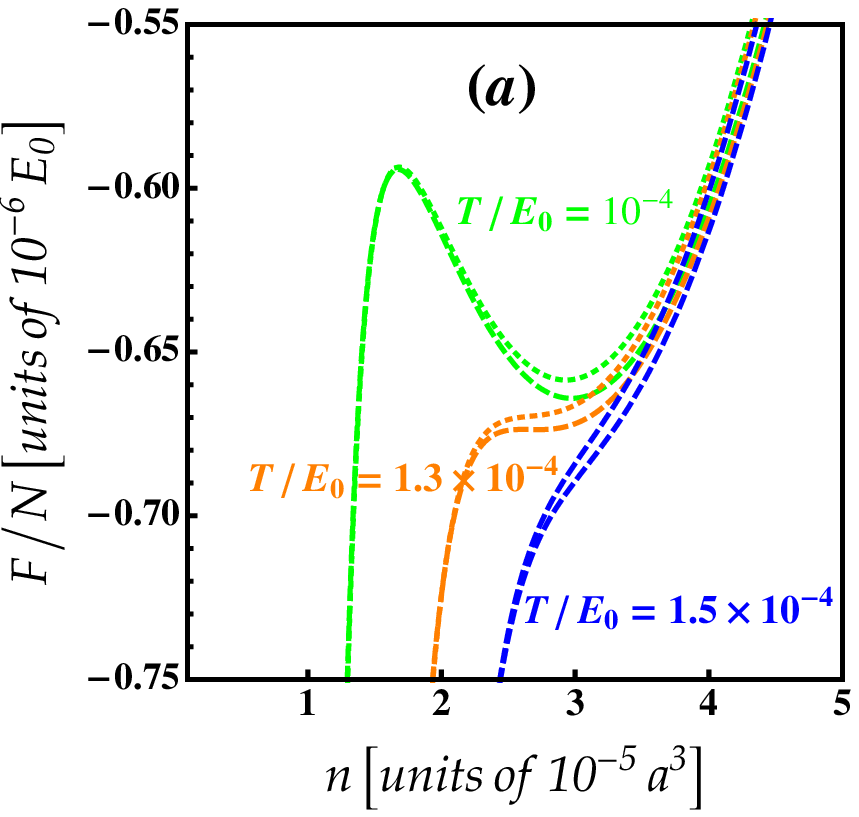}
                \includegraphics[scale=0.45]{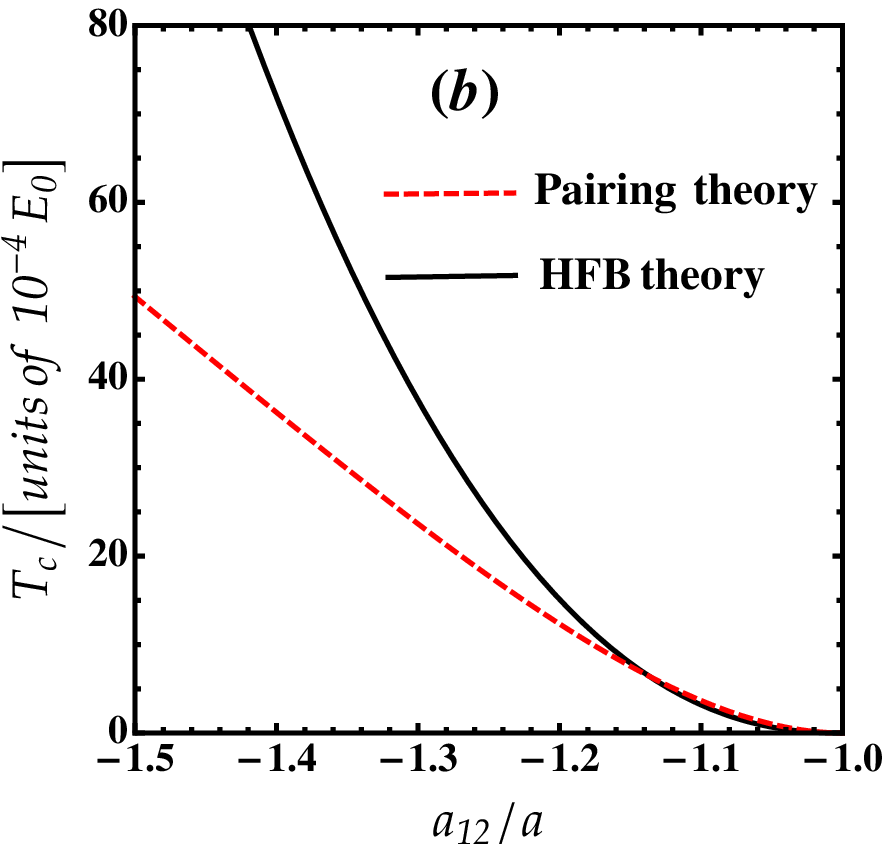}}
		\caption{ (a) Free energy as a function of the density for $a_{12}=-1.1 a$ at different values of temperature.
                               (b) Critical temperature $T_c$ of a droplet as a function of the inter-species interaction strength $a_{12}/a$.
                                 Dotted lines are our results up to first-order in $\tilde m$ and $\tilde n$. Dashed lines are  our results up to the second-order in $\tilde m$ and $\tilde n$.}
\label{Fegy} 
\end{figure}

We now focus ourselves to calculate the noncondensate and anomalous densities in the equilibrium phase.
Inserting Eq.(\ref{eqd}) into Eqs.(\ref{NCD}) and (\ref{MCD}), one can  rewrite the noncondensate and anomalous densities of the droplet in terms of
the equilibrium density up to first-order :
\begin{align} \label{ncdeq}
\tilde{n}_{eq}=\frac{2\sqrt{2}} {3} n_{eq} \sqrt{ \frac{a^3 n_{eq}} {\pi} } \sum_{\pm} \left(\frac{\delta a} {a}\right)_{\pm}^{3/2},
\end{align}
and 
\begin{align} \label{andeq}
\tilde{m}_{eq}=2\sqrt{2} \, n_{eq} \sqrt{ \frac{a^3 n_{eq}} {\pi} } \sum_{\pm} \left(\frac{\delta a} {a}\right)_{\pm}^{3/2}.
\end{align}
To the best of our knowledge,  these quantities have never been determined in the literature.
Remarkably, Eqs.(\ref{ncdeq}) and (\ref{andeq}) show that the anomalous density of the droplet is larger than the noncondensed density specifically for large $|a_{12}/a|$
as in the case of an ordinary BEC.
We see from  Fig.\ref{AND} that $\tilde{n}_{eq}$ and $\tilde{m}_{eq}$ increase with the ratio $|a_{12}/a|$ which is indeed natural since 
both quantities arise from interactions.

\subsection{Self-bound droplet at finite temperature}

At finite temperature, the free energy (\ref{LHY1}) can be rewritten as a function of the small parameter of the theory $na^3\ll1$ as:
\begin{align}\label{LHY3}
\frac{F}{NE_0} =&  \frac{E}{NE_0} -\frac{8 \sqrt{\pi}}{675} (n a^3)^{1/6} \bigg(1+\frac{\tilde m-\tilde n}{n}\bigg)^{-3/2} \nonumber\\
&\times \sum_{\pm} \left(\frac{\delta a}{a}\right)_{\pm}^{-3/2} \left(\frac{T}{E_0 }\right)^4, 
\end{align}
where $E$ is given in Eq.(\ref{LHY2}). 

In Fig.\ref{Fegy}.(a), we present the selfconsistent  solution of the free energy (\ref{LHY3}) up to second order for several values of temperature.
We see that the local minimum disappears at the critical temperature $T_c \simeq 1.3 \times 10^{-4} E_0$ at which 
the thermal fluctuations compensate the repulsive quantum fluctuations. This value is greater by $\sim 23\%$ than the threshold temperature obtained in Refs.\cite{Hu,Ota}
owing to the higher-order beyond-LHY corrections. 
At higher temperatures ($T\simeq 1.5 \times 10^{-4} E_0$), the self-bound droplet is completely evaporated due to the strong thermal fluctuations.
Another important remark is that the higher-order correlations may lead to a shift in the critical temperature.

A useful analytical expression for the critical temperature above which the droplet destabilizes can be extracted by minimizing 
the free energy (\ref{LHY3}) with respect to the density. This yields
\begin{align}  \label{Tempc} 
&\frac{ T_c} {E_0}= 0.0114 \left(\frac{\delta a} {a}\right)_+^2 \bigg[72+25\left(\frac{\delta a} {a}\right)_+\bigg]^{1/4}.
\end{align}
In Fig.\ref{Fegy}.(b) we plot the critical temperature from Eq.(\ref{Tempc}) and compare it with the results of  Ref.\cite{Hu}.
We observe that $T_c$ decreases with decreasing the ratio $|a_{12}/a|$. 
Our findings agree with the pairing predictions only for small $|a_{12}/a|$ and then both theories diverge from each other
on account of the higher-order contribution of the anomalous correlations.

\begin{figure}
		\includegraphics[scale=0.8]{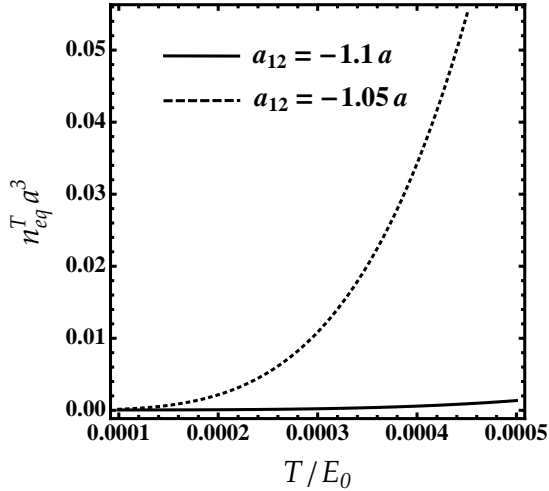}
		\caption{ Thermal equilibrium density as a function of the temperature for different values of interspecies interaction strength.}
\label{Tequi} 
\end{figure}

Treating the temperature term as a perturbation \cite{Ayb}, one gets for the equilibrium density:
			\begin{align}\label{Eq1}
	n_{eq}^T \approx n_{eq} \left[1 + \frac{5 (\delta a/a)_+^{-4}}{128 \sqrt{3}}\Big(\frac{T/E_0}{   a^3 n_{eq} }\Big)^4\right],
	\end{align}
and its behavior is displayed in Fig.\ref{Tequi}. We see that $n_{eq}^{T}$ is increasing with temperature and interspecies interactions.

\section{Ground-state properties} \label{GSP}

In this section we examine the ground-state properties of the droplet by discussing the effects of higher-order quantum and thermal corrections 
in the different ground-state solutions such as the density profiles and the collective modes.

\subsection{Density profiles}

Following the method outlined by Petrov \cite{Petrov}, we rescale the wavefunction of each species as:
$\Phi(r,t)= \sqrt{n_0} \phi(r,t)$, where $\phi (r,t)$ is a scalar wavefunction common to both species. 
It is convenient to introduce the dimensionless parameters : $\tilde{N}= N/(n_0 \xi^3)$, $\tilde{r}=r/\xi$, and $\tilde{t}=t/\tau$, where
$\xi= \sqrt{6 \hbar^2/ \left(m |\delta g | n_0\right)}$, and $\tau = 6 \hbar/\left(|\delta g| n_0\right)$.

\begin{figure}
		\includegraphics[scale=0.44]{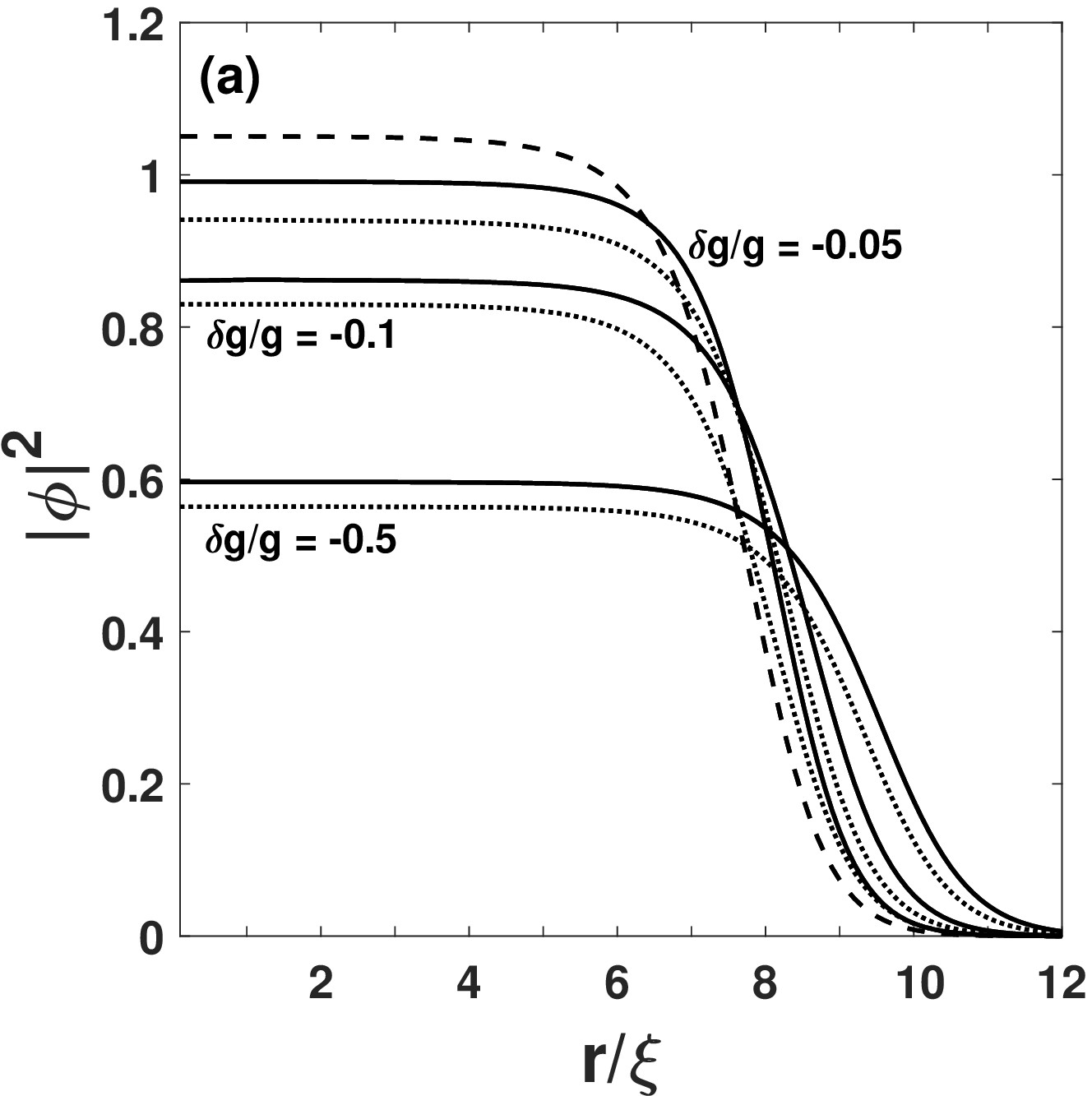}
                \includegraphics[scale=0.44]{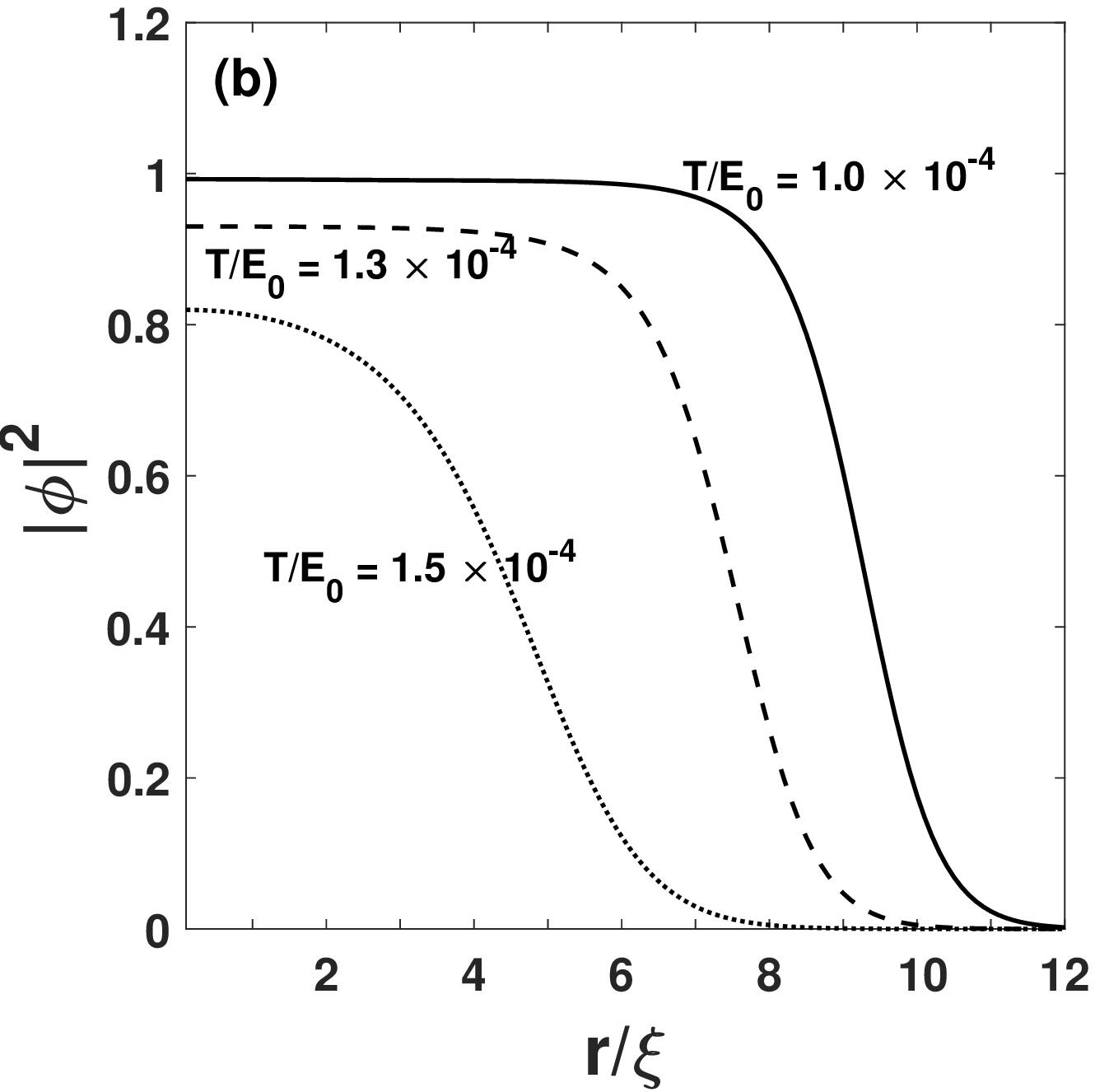}
		\caption{ (a) Density profile of the self-bound droplet obtained from Eq.(\ref{GGP})  at zero temperature for different values of interaction strength $\delta g/g$, 
                and $\tilde N= 3000$.
                  Solid lines: Our results. Dashed lines: Petrov's results \cite{Petrov}. Dotted lines: Findings of Ref.\cite{Ota}.
                 (b) Density profile of the self-bound droplet at different values of temperature for $g_{12} / g =-1.1$. }
\label{Denprof}
\end{figure}

Upon inserting expressions of $\tilde{m}$ and $\tilde{n}$ up to the second-order into Eq.(\ref{egy}), and using our variational TDHFB equation
$i d  \phi/d t =d{\cal E}/d \phi$ \cite{Boudj1,Boudj2}, we get the generalized finite-temperature GPE
\begin{align} \label{GGP}  
&i\frac{d\phi} {d\tilde t}= \bigg \{ -\frac{1}{2} \Delta_{\bf \tilde r} -3|\phi|^2+ \frac{5}{2} [1+\alpha |\phi| (1+ \alpha |\phi|)^{3/2}]^{5/2}|\phi|^3  \nonumber\\
&+\frac{5}{2} \alpha^{\text{T}} |\phi|^{-1}\bigg\}\phi,  
\end{align}
where $\alpha =(4\sqrt{2}/3) (\delta g/g)_+^{3/2}\sqrt{a^3 n_0 /\pi}$, and 
 $\alpha^{\text{T}}= - (4/9 )\sqrt{a^3 n_0 /\pi} (\delta g/g)_+^{-1/2}\big(T/4 a^3 n_0 E_0\big)^2$.
For $\alpha^{\text{T}}=0$, Eq.(\ref{GGP}) reduces to that used in Ref.\cite{Ota}. 
For $\alpha=\alpha^{\text{T}}=0$,  one recovers the Petrov's generalized GPE  \cite{Petrov}.
Equation (\ref{GGP}) is appealing since it obviously contains higher-order quantum corrections and temperature effects.
It is valid when the spatial variations of the condensate density  are small over the extended healing length, $\xi$, validating a description of a locally homogeneous system.

We solve numerically Eq.(\ref{GGP}) and the corresponding results are captured in Fig. \ref{Denprof}.

It is clearly seen in  Fig. \ref{Denprof}.(a) that as the interaction strength $|\delta g/g|$ increases, the central density decreases and the size of the droplet increases
in agreement with the findings of \cite{Ota}. This may provoke a shift in the critical number compared to the universal results of Ref.\cite{Petrov}.

Figure \ref{Denprof} (b) shows that the droplet exhibits a flat-top profile at the equilibrium (i.e. at $T<T_c$).
When the temperature augments, a dramatic change in the shape of the spatial density profile is observed where the condensate is a Gaussian-like shape
at $T\geq T_c$. This signals the evaporation of the droplet to the gas state as is foreseen above.

\subsection{Collective modes}

The collective modes of the self-bound droplet can be analytically explored employing a Gaussian ansatz
\begin{eqnarray}\label{Eq:phi}
\phi(r)=A \exp\bigg[-\left(\frac{1}{2 \sigma^2}+i \beta \right) r^2\bigg],
\end{eqnarray}
where  $A =\sqrt{\tilde N/\left(\pi^{3/2}\sigma^3\right)}$ is the normalization factor, $\sigma$,  and $\beta$ are the variational parameters 
standing for the width and  the phase of the  condensate, respectively.

\begin{figure}[hp]
		\includegraphics[scale=0.8]{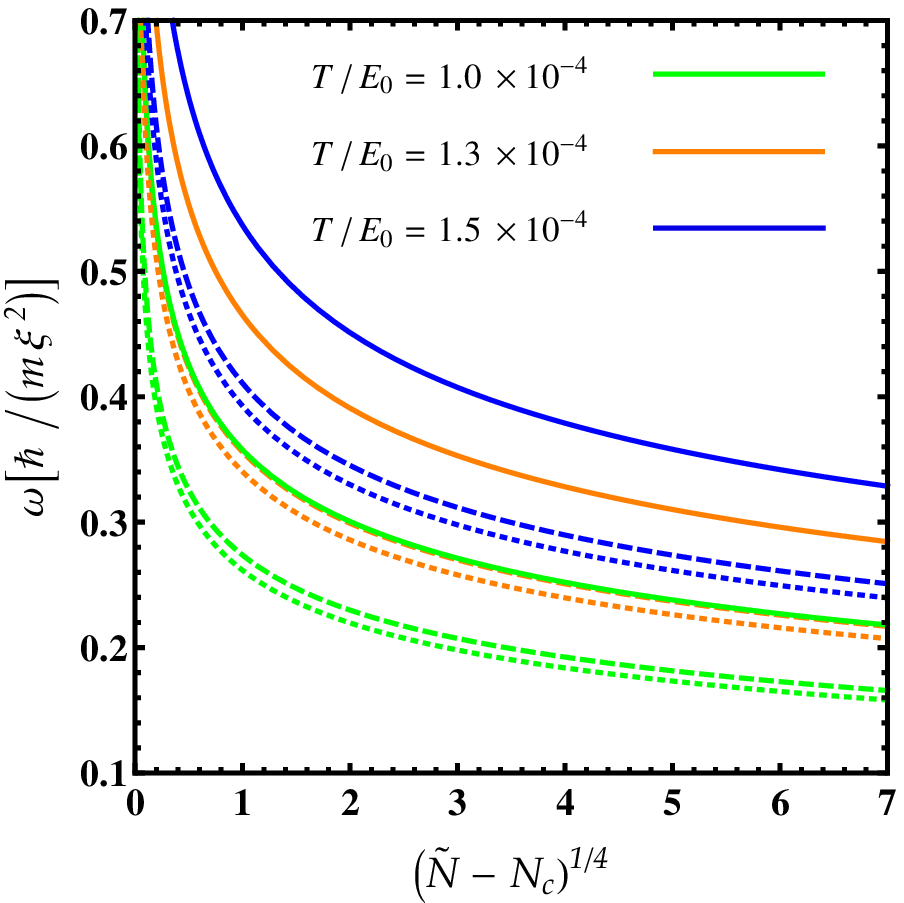}
		\caption{Frequencies of low-lying collective excitation modes  as a function of the total atom number
			in units of $\tilde{N}$. Dotted line: $\delta g/g = -0.05$, dashed line: $\delta g/g = -0.1$, solid line: $\delta g/g = -0.5$.
Here $N_c=18.65$ is the critical number found in Ref.\cite{Petrov}. }
\label{breathing}
\end{figure}

The dynamics of the droplet width can be obtained from the Euler-Lagrange equations
	\begin{equation}\label{Eq:Va4}
	\ddot{\sigma}=-\frac{d V_{eff}(\sigma)}{d \sigma},
	\end{equation}
where the effective potential is given by:
	\begin{align}\label{Eq:Veff}
		V_{eff}(\sigma)&=-\frac{\tilde{N}}{2 \sigma^2}+\frac{\sqrt{2}}{3 \pi^{3/2}}\frac{\tilde{N}}{\sigma^3}
	+\frac{8\sqrt{2/5}}{27 \pi^{9/4}}\frac{\tilde{N}^{3/2}}{\sigma^{9/2} }\\
	&+\frac{4 \sqrt{2/5}}{18 \pi^3}\frac{\tilde{N}}{\sigma^6} \alpha+\frac{16 \sqrt{2/\tilde{N}} \pi^{3/4}\sigma }{9} \alpha^T.\nonumber
	\end{align}
Frequencies of the low-lying collective modes are determined by linearizing Eq.(\ref{Eq:Veff})  around the equilibrium solutions. 
The breathing modes can be found by inserting the decomposition $\sigma (t)=\sigma+\delta\sigma(t) $ 
(with $\delta\sigma(t)=\delta\sigma \, e^{i \omega t}$ and $\delta\sigma\ll \sigma$) into Eq.(\ref{Eq:Veff}) 
and expanding the effective potential into a Taylor series. As shown in Fig.\ref{breathing}, 
both temperature and interactions $|\delta g/g|$ tend to increase the frequency of the breathing modes. 
Another important remark is that higher-order effects in the breathing modes are substantial.

\section{Conclusion and outlook} \label{concl}

In this work we investigated the effects of higher-order quantum and thermal fluctuations on the properties of a symmetric mixture self-bound droplet using
the beyond-LHY theory.
We derived useful formulas for the energy, the equilibrium, the normal and anomalous densities, and the critical temperature of the self-bound droplet.
Our predictions have been tested against DMC data and the existing theoretical results, and  excellent agreement is found.
Finite-temperature effects on the ground-state properties of the self-bound droplet have been also addressed by numerically solving the extended finite-temperature GPE.
We showed that the inclusion of higher-order quantum and thermal corrections may shift the density profiles and frequencies of the breathing oscillations. 
We conjecture that our model can be extended to the asymmetric droplet for a direct comparison with recent experiments. 
This is one of our future works.


\section{Acknowledgments}

We are grateful to Jordi Boronat and Viktor Cikojevi\'c for giving us the DMC data.
We acknowledge support from the Algerian ministry of higher education and scientific research under Research Grant No. PRFU-B00L02UN020120190001.

\end{document}